\documentclass[english,aps,secnumarabic,amsmath,twocolumn,amssymbs,showpacs]{revtex4}
\usepackage[T1]{fontenc}
\usepackage[latin9]{inputenc}
\usepackage[active]{srcltx}
\usepackage{amssymb}
\usepackage{graphicx}

\makeatletter
\@ifundefined{textcolor}{}
{%
 \definecolor{BLACK}{gray}{0}
 \definecolor{WHITE}{gray}{1}
 \definecolor{RED}{rgb}{1,0,0}
 \definecolor{GREEN}{rgb}{0,1,0}
 \definecolor{BLUE}{rgb}{0,0,1}
 \definecolor{CYAN}{cmyk}{1,0,0,0}
 \definecolor{MAGENTA}{cmyk}{0,1,0,0}
 \definecolor{YELLOW}{cmyk}{0,0,1,0}
}
\usepackage{setspace}

\usepackage{babel}

\begin{document}

\title{Measuring topological invariants in photonic systems}
\author{M. Hafezi}
\affiliation{Joint Quantum Institute, NIST/University of Maryland, College Park
MD}

\begin{abstract}
Motivated by the recent theoretical and experimental progress in implementing topological orders with photons, we analyze photonic systems with different topologies and present
a scheme to probe their topological features. Specifically,  we propose a scheme to modify the boundary phases  to manipulate edge state dynamics. Such a scheme allows one to measure the winding number of the edge states. Furthermore, we discuss the effect of loss and disorder on the validity of our approach.
\end{abstract}
\maketitle

Topology plays a fundamental role in many physical phenomena in two-dimensional systems. Most famous examples are  various quantum Hall effects in electronic systems \cite{Klitzing:1980,tsui82,Konig:2007}. Recently,  there has been a surge of interest in studying topological orders in non-electronic systems, ranging from ultra cold atoms  to photons. In atomic systems, there are ongoing efforts to synthesize gauge fields in many groups \cite{Dalibard:2011p46769,Lin:2009p28223,Struck:2012gc,Aidelsburger:2011hl}.  More
recently, optical systems have been under investigation to implement gauge fields using various schemes: application of strong
magnetic field \cite{Haldane:2008,Wang:2009p27234}, polarization
scheme \cite{Umucalilar:2011}, opto-mechanics \cite{Hafezi:2011ui},
differential optical paths \cite{Hafezi:2011delay,Hafezi:2013ub,Liang:2012ty}, bi-anisotropic
metamaterials \cite{Khanikaev:2012uf}, harmonic modulation \cite{Fan:vy}, side-coupled waveguide systems \cite{Kraus:2012,Verbin:2013bv}, strain induced
magnetic field \cite{Rechtsman:2012tc} and helical waveguides \cite{Rechtsman:2013fe}.

While the implementation of gauge fields has been achieved, the detection of the expected topological orders remains elusive, due to the inapplicability of the conventional Hall conductance measurements to atomic and photonic systems. There have been various proposals to detect topological order in atomic  \cite{Price:2011vi,Alba:2011p46677,Abanin:2012uj,Goldman:2012fd,PatrickLee:2013} and recently photons systems \cite{Ozawa:2013tt,Longhi:2013cx}, which are generally based on manipulation of Bloch states.  However, the following  question has not yet been  addressed: how can one measure the integer topological invariants, e.g.  the winding number of the edge states or the Chern number of the bulk state in a photonic system? In particular, how do the integer values manifest themselves in an optical realization of quantum Hall Hamiltonians. In this Letter, we propose a scheme to measure the integer topological invariants of a photonic system, by manipulating the boundary conditions. We benefit from individual site addressability to manipulate the synthetic gauge field at the boundary -- a property which is difficult to achieve in electronic and atomic systems. 

The main idea of our approach relies on the ability to introduce a nonzero phase in the boundary conditions. Such phase is equivalent to a magnetic flux  threading the holes of the system, when the system manifold is not simply connected. If the system has an edge state around that hole, the insertion of the magnetic flux shifts the momentum of that edge state. Once an entire magnetic flux quantum threads the hole, the edge state spectrum should return to its original form, while an integer number of edge states have shifted during this process. This integer number is the winding number of the edge state. We show that, in a photonic implementation, such a spectral shift and edge state transfer can be experimentally observed, using standard transmission spectroscopy. We note that our proposal could be applied to all topologically-ordered photonic systems, ranging from radio-frequency \cite{Jia:2013tx} and microwave  \cite{Shanks:2013hi} to optical  domain. For concreteness, we focus on the optical scheme based on a platform proposed in Ref. \cite{Hafezi:2011delay} and recently implemented in Ref.\cite{Hafezi:2013ub}, using Silicon-on-insulator technology.

\begin{figure}[t]
\includegraphics[width=0.4\textwidth]{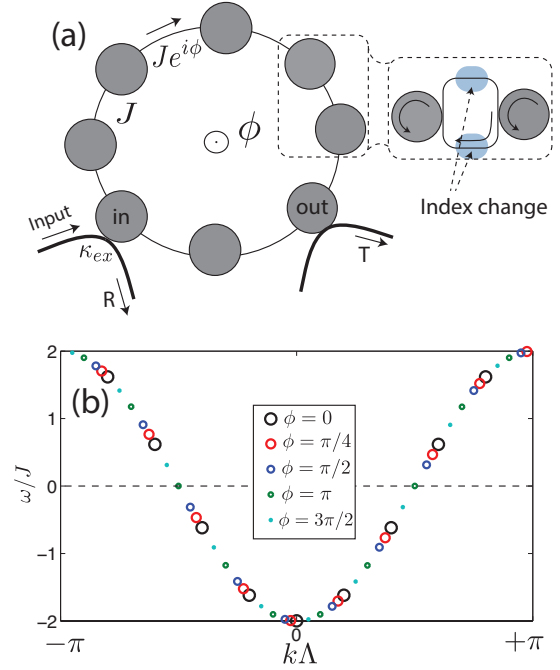}
\caption{ (a) Coupled resonators in a ring configuration: photons hop between resonators
with the rate $J$, except for one link where they hop with $Je^{i\phi}$. The inset shows that the boundary can be modified by changing the index of refraction of the connecting resonators. (b) Dispersion relation when the boundary phase changes from zero to $2\pi$, for a ring of 10 resonators.\label{fig:ring-topology}:}
\end{figure}

\emph{Ring.} 
We start with the simplest topology which is that of a ring. We consider
an array of  $N$ coupled resonators that form a ring and study the effect
of a synthetic magnetic flux  threading the ring, as shown in
Fig.\ref{fig:ring-topology}. The Hamiltonian of the system is given by:

\begin{equation}
H_{ring}=-J \sum_{i=1}^{N-1} \hat{a}^\dagger_i \hat{a}_{i+1} +h.c.
\end{equation}
where $J$ is the tunneling rate between two sites and $\hat{a}^\dagger_i$ is the creation operator at the site $i$. The Hamiltonian terms that describes the coupling between the first and the last site is: $-J \hat{a}^\dagger_1 \hat{a}_N e^{i \phi}+h.c.$, where $\phi$ is the tunneling phase. In other words, the twist angle in the generalized boundary condition, can be generated by such a term in the system Hamiltonian. For charged particles, this phase can be obtained by introducing a magnetic flux in the middle of the ring, whereas in our system, this phase has to be artificially engineered.  Note that such phase does not have to be local on the last link, i.e., it can be distributed around the loop and generate the same effect.

Following Ref.\cite{Hafezi:2011delay}, the tunneling between the resonators can be induced by connecting loops that are anti-resonant
with the original resonators, as shown in Fig.1 inset. Specifically, we assume that the perimeter of the resonators is $ m\lambda$ where $m$ is an integer and $\lambda$ is the resonant wavelength. If the perimeter of the connecting loop is chosen to be $m\lambda+3\lambda/2$, it will induce a coupling between two resonators described by the Hamiltonian:  $-J \hat{a}^\dagger_i \hat{a}_j+h.c.$ Now, the tunneling term can take a phase, if the index of refraction of the upper
and lower arms are changed with opposite signs, so that the overall
connecting loop remains anti-resonant, while the forward and backward hopping acquire opposite phases, described by the Hamiltonian: $-J \hat{a}^\dagger_i \hat{a}_j e^{i \phi}+h.c.$.  Such index change can be achieved through optical \cite{Vlasov:2008} or electrical \cite{Xu:2005p29862} carrier injection or thermal tuning \cite{Melloni:10p45138}. Alternatively, a non-reciprocal phase can be induced by modulating the connecting waveguides \cite{Fan:2012ur} or using opto-mechanics \cite{Hafezi:2011ui}. Regardless of the experimental scheme, the dispersion relation of  the ring is: 

\begin{equation}
\omega=-2J\cos(k\Lambda+\phi/N).
\end{equation}
Note that the phase is divided by $N$ since the hopping phase is introduced only at one link and it does not depend on how the phase is distributed over the lattice, as long as the total hopping phase is equal to $\phi$. 
In the context of the conventional tight-binding
model, $k$ can be interpreted as the Bloch wave number and $\Lambda$
as the lattice spacing. Here, $k\Lambda$ is simply the phase difference between two adjacent resonators. In a finite system, the eigen energies are positioned on a finite number of points on the same dispersion curve, as shown in Fig. ~1b.
Changing the twist angle $\phi$ shifts the energy spectrum in one direction, along the dispersion curve. When one flux quantum is inserted ($\phi=0\rightarrow2\pi$), the energy spectrum returns to its original
form, shifting one state in the Brillouin zone, as shown in Fig.~1b. 

\begin{figure}
\center
\includegraphics[width=0.5\textwidth]{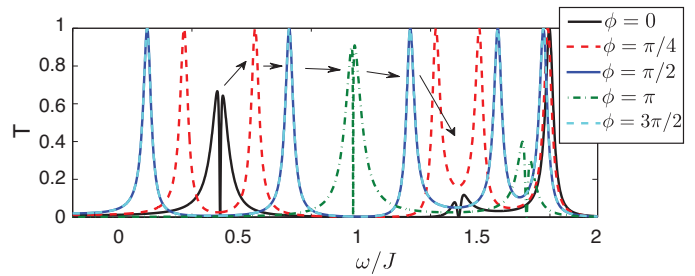}\caption{Probing the state transfer in a ring using transmission spectroscopy. The simulation is performed for 10 sites and $\kappa_{ex}/J=0.1$. The input and output probes are separated by two sites, as shown in Fig.~1b. \label{fig:probing-state-transfer} }
\end{figure}

In a photonic system, such a state transfer can be probed through transmission spectroscopy.  Using the input-output formalism \cite{Gardiner:1985}, the field dynamics of the resonators is given by:
\begin{eqnarray}
\dot{\hat{a}}_{j}&=&i[H,\hat{a}_{j}]-(\delta_{j,\rm{in}}+\delta_{j,\rm{out}})\kappa_{ex}\hat{a}_{j}\nonumber\\
&-&\delta_{j,\rm{in}}\sqrt{2\kappa_{ex}}\mathcal{E}_{in}e^{-i\omega t}
\end{eqnarray}
where $\kappa_{ex}$ is the extrinsic coupling rate between the probing-waveguide
and the resonators. ``in'' (``out'') indices represents the resonators to
which the input (output) probing waveguides are connected. In the linear regime where $\langle\hat{a}_{i}\rangle=a_{i}$, we can obtain the transmission in the output channel as $T=|a_{out}/\mathcal{E}_{in}|^{2}$. Fig.~\ref{fig:probing-state-transfer} shows the transmission spectrum of the system when the array is probed using an input and output waveguide  (shown in Fig.~1a).  

In the absence of the magnetic flux ($\phi=0$), the spectrum is two-fold
degenerate, which corresponds to Bloch waves going  clock-wise and counter-clockwise around the ring.  However, in the presence of the magnetic flux ($\phi \neq 0$), the spectrum is
not necessarily degenerate, and therefore, all the states can be resolved using transmission
spectroscopy in the under-coupled limit ($\kappa_{ex}<4J/N$), i.e., the finite size of the system allows us to track the transfer of the states. We readily observe that when $\phi: 0 \rightarrow 2\pi$, the transmission spectrum returns to its original profile and each peak moves and replaces its adjacent peak. We use similar tool to investigate a two-dimensional system.

\emph{Annulus.} 
We consider an annulus similar to Laughlin-Halperin's argument
\cite{Laughlin:1981p37080,Halperin:1982} to study edge state transfer. Specifically, following Hatsugai's work
\cite{Hatsugai:1993PRB}, we consider a 2D lattice with uniform perpendicular magnetic field, where the Hamiltonian of the system is given by:

\begin{eqnarray}
H_{{\rm mag}} & = & -J\sum_{x,y}\hat{a}_{x+1,y}^{\dagger}\hat{a}_{x,y}e^{i2\pi\alpha y}+\hat{a}_{x,y}^{\dagger}\hat{a}_{x+1,y}e^{-i2\pi\alpha y}\nonumber \\
 &  & \ +\ \hat{a}_{x,y+1}^{\dagger}\hat{a}_{x,y}+\hat{a}_{x,y}^{\dagger}\hat{a}_{x,y+1},\label{eq:mag_hamiltonian}
\end{eqnarray}
where $\hat{a}_{x,y}^{\dagger}$ is the creation operator at the site $(x,y)$, and $\alpha$
characterizes the phase imbalance. Specifically, a photon hopping
around a plaquette, in the clockwise direction, acquires the
phase $2\pi\alpha$, in direct analogy to Aharanov-Bohm phase. Therefore,
$\alpha$ is the effective magnetic flux per plaquette, and therefore, the total
magnetic flux is $N_{\alpha}=\alpha N_{x}N_{y}$. This Hamiltonian has been theoretically proposed in Ref. \cite{Hafezi:2011delay} and experimentally demonstrated in Ref. \cite{Hafezi:2013ub}. 

\begin{figure}[t]
\includegraphics[width=0.45\textwidth]{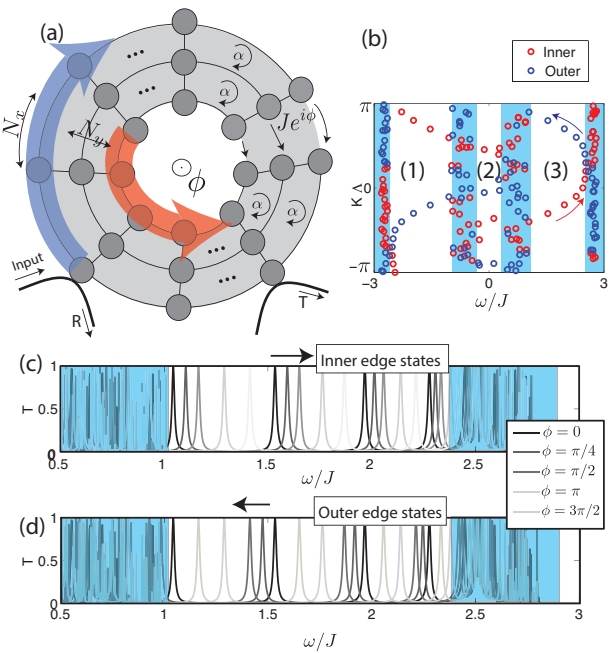}
\caption{ (a) Cylinder/annulus  topology, the system is under a uniform magnetic flux $\alpha$ and the hole is threaded with a magnetic flux $\phi$. (b) shows the dispersion relation for $\phi=0$. The bulk gap indices are shown. (c,d) The state transfer for inner and outer edges, respectively, where the threading  flux is changed from zero to $2\pi$. The simulation is performed for $(N_x,Ny)=(20,10)$, $\alpha=1/4$ and  $\kappa_{ex}/J=0.1$. Blue shades represent the bulk states.}
\end{figure}

For an infinite system, the Hamiltonian of Eq.(2) yields Hofstadter butterfly spectrum \cite{hofstadter}. In particular, when the magnetic flux is rational,  $\alpha=p/q$ with mutually prime integers,  the system has $q$ distinct (gapped) bands \cite{hofstadter}. On a finite annulus, or any other equivalent topology, e.g. a cylinder or a square with a hole in the middle, the system havs edge states which are spectrally located between the magnetic bands and are spatially confined at the edges.  The dispersion of the edge states is shown in Fig.~2b. As shown by Hatsugai  \cite{Hatsugai:1993PRB}, the winding number of the such edge state, which is related to the Chern number of the bulk states,  and is given by the following Diophantine equation:
\begin{equation}
\label{eq:diophantine}
n=s_nq+t_np~~,~~|t_n|\leq q/2 ,
\end{equation}
where $t_n$ and $s_n$ are integers and $n$ is the gap index ($1\leq n \leq q-1$), and $t_n$ is the winding number of the $n-$th gap.

Besides the overall uniform magnetic flux ($\alpha$), we assume that the system is threaded with a synthetic magnetic flux $\phi$ through the hole of the annulus.  
When the magnetic flux is changed from zero to one, the edge states are transferred. We can easily trace
the edge states and count how many of them have been transferred during
the insertion of a magnetic flux. By connecting the probing waveguides
to outer (inner) edges, we can selectively, couple to outer (inner)
edge states, respectively, as shown in Fig.3c-d. In particular, when
the connecting waveguides are coupled to the outer (inner) edge of
the system, the coupling to the inner (outer) edge is exponentially suppressed as $\exp(-Ny/l_B)$ where $l_B^{-1}=\sqrt{2\pi\alpha}$ is the magnetic length. We focus on the third gap, where for a system with $\alpha=1/4$, the winding number is one ($t_3=-1$), according to Eq.(\ref{eq:diophantine}). As shown in Fig.~3c-d, the edge states are transferred by one peak, in agreement with the value of the winding number. Note that the inner and outer edge spectrum move in opposite directions.
 
 In Fig.~4, we consider another situation where $\alpha=1/6$.  For the fourth gap (i.e., $0.6\leq\omega\leq1.4$), the winding number is $t_4=-2$. Therefore, the edge state resonances shift by two peaks, when the system hole is threaded with a magnetic flux quantum, as shown in Fig.~4b. In general if the winding number of the edge state is $t$, the edge spectrum shifts by $t$ peaks, when the twist angle is change by $2\pi$. Moreover, the moving direction of the peak corresponds to the sign of the winding number of the edge states.

\begin{figure}[h]
\includegraphics[width=0.45\textwidth]{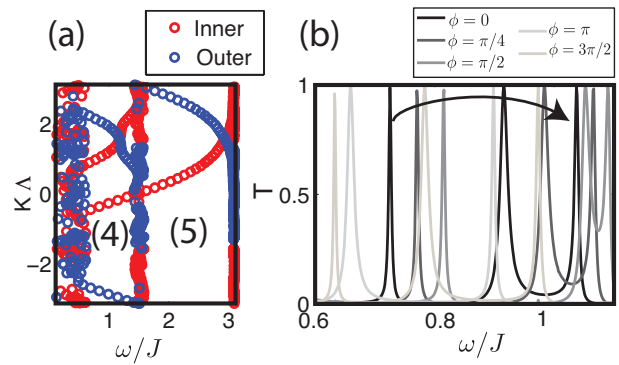}
\caption{ (a) Dispersion relation of the fourth and the fifth bands for $\alpha=1/6$ for a 70x20  lattice. (b) Transmission spectrum as a function of the twist angle $\phi$  for a 20x14 lattice. }
\end{figure}

\emph{Effect of disorder and loss.} In an experimental realization, photonic systems are impaired by loss and disorders.  Therefore, we evaluate the effect of such errors and show that the proposed scheme can still probe the topological invariants. The major source of loss in silicon-on-insulator is the propagation loss in Silicon rings, where the guided photons leave the waveguides through elastic scattering \cite{Vlasov:2004}. We characterize such loss in our Hamiltonian, in form of $-i\kappa_{in} \hat{a}^\dagger_i \hat{a}_i$  where $\kappa_{in}$ is the field intrinsic decay rate to undesired modes. Fig.~5a shows the effect of such loss on the system of Fig.~2. We observe that the presence of loss decreases the contrast of the transmission spectrum, and as long as the loss rate is at most an order of magnitude lower than the tunneling rate, the transmission peaks are quite discernible. The other source of error is the frequency mismatch between neighboring resonators. Such disorder, which is caused by the width  and height variation of the waveguides,  is a common problem in integrated photonics \cite{Barwicz:2006,Xia:2007,Ferrari:2009} and can be characterized by a random on-site potential at each site $U_{i} \hat{a}^\dagger_i \hat{a}_i$ \cite{Hafezi:2011delay}. Fig.~5b shows the effect of such disorder on the transmission spectrum. The grey area highlights one standard deviation from the averaged transmission in the presence of non-magnetic disorder evaluated for a hundred realizations. As expected from the theory of integer quantum Hall effect, disorder leads to broadening of the edge state resonances, however, they are still resolvable, as shown in Fig.~5b. Since the bulk states are more susceptible to disorder, the corresponding resonances are washed out, as shown on the left and right of the spectrum. In contrast, the edge states are less susceptible  to disorder, and their spectrum is robust.

\begin{figure}[t]
\includegraphics[width=0.48\textwidth]{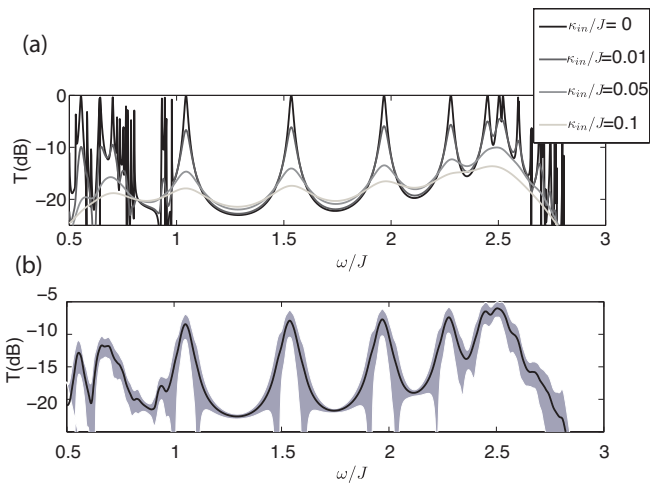}
\caption{(a) Transmission spectrum for different values of intrinsic loss $\kappa_{in}/J=0,0.01,0.05,0.1$  (b) Transmission spectrum in the presence of intrinsic loss ($\kappa_{in}=0.01$) and a gaussian disorder  ($\sigma(U)/J=0.1$). All the other parameters are the same as Fig.~3. }
\end{figure}

In conclusion, we have shown that different topologies can be implemented in photonic system and their  integer topological invariants can be measured, using standard transmission spectroscopy. The focus of this Letter was the linear regime. Recently,  it has been shown that, in the presence of strong optical nonlinearity, an externally driven system can form fractional quantum Hall states, such as Laughlin states \cite{Umucalilar:2011b,Hafezi:2013wh}. An interesting research direction is to extend these ideas to investigate topological invariants in such driven interacting systems.

\emph{Acknowledgements.} We thank Y. Hatsugai, P. Zoller, I. Carusotto, N. Goldman, B. Halperin, J. Taylor, O. Zilberberg and J. Peters  for fruitful discussions. This research was supported by the ARO MURI award W911NF0910406, and the NSF through the Physics Frontier Center at the Joint Quantum Institute.

\bibliographystyle{apsrev}

\end{document}